\documentstyle[12pt]{article}
\newcommand {\Pom}  {I\hspace{-0.3em}P}
\def\lsim{\mathrel{\rlap{\lower4pt\hbox{\hskip1pt$\sim$}}
    \raise1pt\hbox{$<$}}}         
\def\gsim{\mathrel{\rlap{\lower4pt\hbox{\hskip1pt$\sim$}}
    \raise1pt\hbox{$>$}}}         

\begin{document}

\vspace*{1.5cm}
\begin{LARGE}
\begin{bf}
\begin{center}
A Spectrometer to Study Elastic and Diffractive Physics at LHC\\
\end{center}
\end{bf}
\end{LARGE}
\begin{LARGE}
\begin{center}
M.Arneodo, S.Maselli, C.Peroni \\
INFN and University of Torino \\[8pt]
\today
\end{center}
\end{LARGE}

\vspace{2.0cm}
\indent
{\bf Abstract}

\vspace{0.5cm}
\indent

The possibility to study
elastic and diffractive physics in $pp$ collisions at LHC is discussed.
For this purpose we have considered detectors close to the beam in conjunction
with the magnetic elements of the accelerator to
provide a high precision spectrometer for very forward final
state protons.
The geometrical acceptance is given and the momentum resolution is
calculated for different spatial resolution detectors.

\vspace{1.cm}

\indent
\section{\bf Introduction }

\vspace{0.5cm}
\indent

Elastic and diffractive scattering processes 
are generally interpreted in terms of Pomeron ($\Pom$) exchange 
\cite{Pom}.
They are characterized by a final state which contains particles
emitted at very small angles with respect to the incoming beam and
carry a substantial fraction of the available energy.
At collider experiments these particles escape undetected down the beam pipe.
The Roman pot technique \cite{pots} allows to position detectors directly
in the vacuum pipe at a very small distance from the beam.

Since 1960 elastic and diffractive scattering processes 
have been studied in hadron-hadron interactions at CERN \cite{hadhad} 
and elsewhere.
Diffractive events in deep inelastic $ep$ scattering have been also measured 
at HERA \cite{HERA}. 
Within the ZEUS experiment, the Leading Proton Spectromenter (LPS)
\cite{LPS}
has been built to measure the protons scattered
at very small angle with respect to the incoming beam direction.

\begin{figure}[p]
\begin{picture}(450,550)(0,100)
\includegraphics{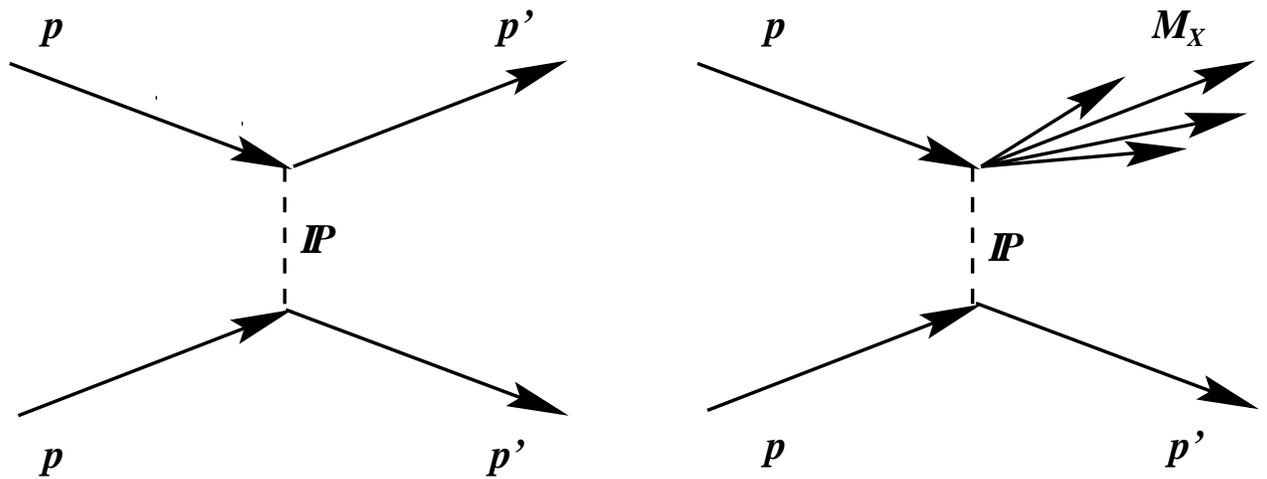}
\end{picture}
\vspace{-3.cm}
\caption{Single pomeron ($\Pom$) exchange in elastic 
(a) and single diffractive (b) $pp$ interactions. 
In interactions of type (b) a final hadronic system $M_X$
is produced.}
\label{Fig:fig1}
\end{figure}

In this paper we investigate the possibility to study elastic and diffractive
events in $pp$ collisions at LHC energy ($\sqrt s =$ 14 TeV).
In particular we study
single diffractive events characterized by one of the protons
emerging from the interaction isolated in rapidity and with its momentum
very close to that of the beam while the other diffracts
into a state of mass $M_X$.
The Feynman graphs of the interaction are shown 
schematically in  Fig.\ref{Fig:fig1}.
Preliminary studies on this process are presented in \cite{diffLHC}.
A full acceptance detector 
could be built in the last free 
LHC intersection region (IP4) where one could have low luminosities,
of the order of $10^{30}$ cm$^{-2}$ s$^{-1}$.
An experimental apparatus
for the complete forward region
has been discussed in  
\cite{Zichi}, \cite{felix1} and \cite{felix2}.

In the following we present the simulation of 
spectrometers in the very forward region 
along one of the two outgoing proton beam arms 
at a sufficient distance from the interaction point
in order to accept final state particles produced at very small angle
with respect to the incoming proton beam direction.
The present paper is an extended and improved version of
\cite{silvia}.

\vspace{1.cm}

\indent
\section{\bf Simulation of elastic and diffractive $\boldmath pp$ events }

\vspace{0.5cm}
\indent

The process under study is the reaction
$pp \rightarrow pX$ (Fig.\ref{Fig:fig1}) 
in which one of the protons remains intact and loses
a small fraction of its momentum $1-x_{L}$, 
with $x_{L} = {|p'|\over{|p|}},$ where $p$ and $p'$
are the incoming and outgoing proton momenta, respectively.
This quantity is related to the mass $M_X$ of the hadronic
system produced in the interaction by the relation:
$$\xi = {1-x_{L}} = {{M^{2}_{X}}\over{s}},$$
where $s$ is the square of total centre of mass energy.
Elastic scattering is characterized by $x_{L} = 1$ while diffractive 
processes
dominate for $x_{L} > 0.9$.
The transverse momentum  $p_{T}$  of the scattered proton
is generally small: $p_{T} \leq 1\ \mbox{GeV/c}$.

In order to select the possible regions along the proton beam line
where these processes can be studied, we need to examine in detail
the properties of the proton beam and the geometrical restrictions
imposed by the vacuum pipe. 

Two possible ways to study the beam profile as it proceeds through the 
magnetic elements are summarized in the following.
The first method makes use of the beta function $\beta$ and the phase 
advance 
$\Delta{\mu}$ parameters extracted 
directly from the LHC MAD tables \cite{MAD}.
The transverse distance in the bending plane 
of a particle from the nominal beam, $x$, and 
its angle $\theta_{x}$ with respect to the nominal beam direction can be 
expressed as:
   \begin{equation}
       x(l)\; =\; \sqrt {{\beta_x}(l)\over{\beta_{x}(0)}}\,cos{\Delta{\mu}}\;
              x(0)\;+\;\sqrt{{\beta_x}(l)\,\beta_{x}(0)}\,
              sin{\Delta{\mu}}\;
              \theta_{x}(0)\;+\;\xi\;D_{x} ,
          \label{bpara1}
   \end{equation}
and
   \begin{equation}
\theta_{x}(l)={-1 \over {\sqrt{{\beta_x}(l) \beta_{x}(0)}}}
  (sin{\Delta{\mu}}+\alpha_x(l) cos{\Delta{\mu}})x(0)+
\sqrt{{\beta_x}(0)\over{\beta_{x}(l)}}(cos{\Delta{\mu}}-
\alpha_x(l) sin{\Delta{\mu}})\theta_{x}(0)+\xi D_{x}'
          \label{bpara2},
   \end{equation}
where 
   \begin{equation}
         \alpha_{x}(l) \; = \; -0.5\;{{d\beta_x} \over {dl}}
   \end{equation}
and $D$ is the dispersion. In the case of elastic scattering $\xi$ is the
natural beam momentum spread of $10^{-4}$.
 
\noindent
The position and angle of the particle at the interaction point are 
indicated as $x(0)$, $\theta_{x}(0).$
Here $l$ is the coordinate along the beam path.
Similar equations can be written for the $y$ plane.
Results using this method are given in \cite{felix1} and \cite{felix2}.

The second way to look at the beam profile is to use
the transport matrices of the beam line: 
              \begin{eqnarray}
      \left( \begin{array}{c}
                     x(l)  \\ \theta_{x}(l)
             \end{array} \right) =
             \left( \begin{array}{cc}
                          H_{11}~~H_{12} \\H_{21}~~H_{22}
                    \end{array}  \right)
             \left( \begin{array}{c}
                        x(0)  \\ \theta_{x}(0)
                    \end{array}  \right),
          \label{matrix}
          \end{eqnarray}
\noindent
where the matrix elements $H_{ij}$ are functions of $x_L$ only.
\par
\begin{figure}[p]
\begin{picture}(450,550)(0,100)
\includegraphics{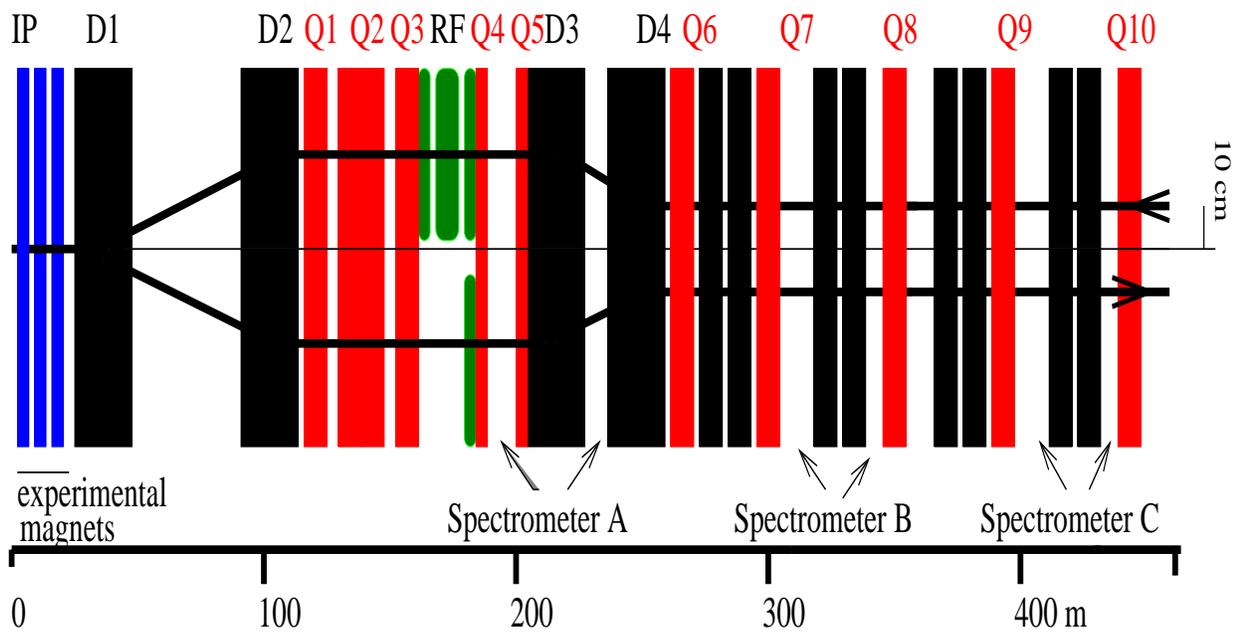}
\end{picture}
\vspace{-2.cm}
\caption{Plan view of a possible insertion in IP4
showing the positions of the three spectrometers
considered in this paper (from [8]).}
\label{Fig:fig2}
\end{figure}

Fig.\ref{Fig:fig2} shows a possible layout for an insertion in 
intersection
region IP4 at LHC which is consistent with the constraints imposed
by the experimental magnets and RF cavities \cite{felix1} and \cite{beamline}.
The experimental equipment for the detection of very forward 
scattered protons
may be located in Roman pots inside the vacuum beam pipe.
Tables \ref{geo1},\ref{geo2},\ref{geo3} 
list the main features of
the magnetic elements of one of the outgoing arms,
from the interaction point up to 450 m.

\begin{figure}[p]
\begin{picture}(450,550)(0,100)
\mbox{}
\vspace{3.0cm}
\includegraphics{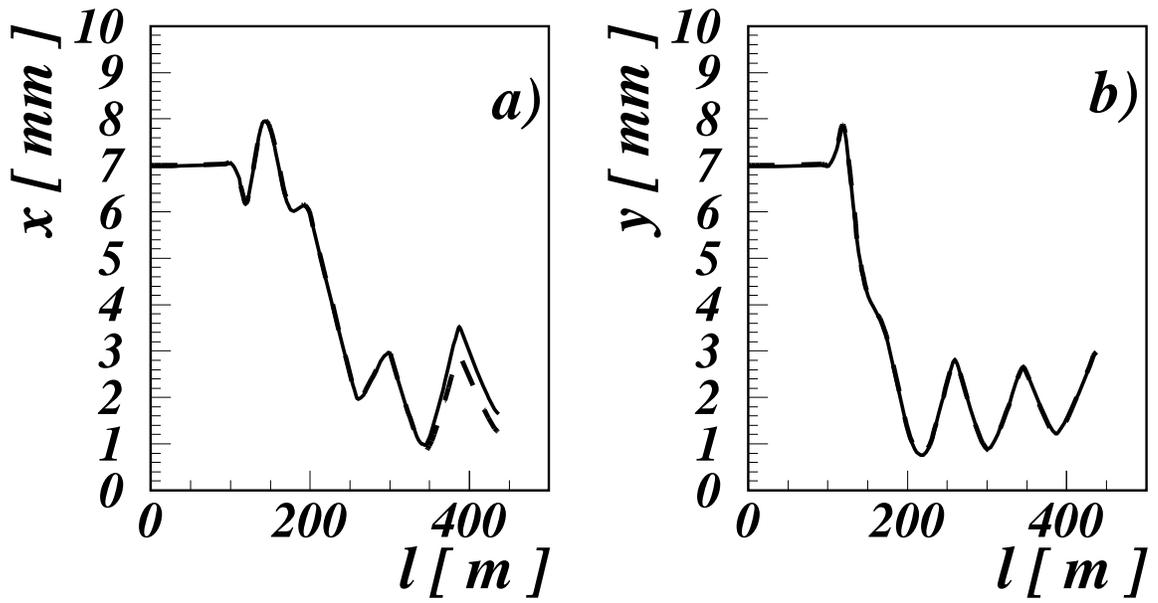}
\end{picture}
\vspace{-3.cm}
\caption{$10 \sigma$ profiles  a) horizontal; b) vertical.
The continuous lines show the orbits calculated with
eqs.(1),(2) (as in [8]) the dashed lines those calculated
with eq.(4). }
\label{Fig:fig3}
\end{figure}

Fig.\ref{Fig:fig3} shows the $10\sigma $ profiles of the beam in the horizontal
and vertical planes. They have been calculated using both  methods
discussed earlier.The orbits agree all along the beam line,
showing that the tracking methods
are equivalent.

All the results presented in this paper
have been obtained with a program which simulates
particle trajectories using the beam optics transport equation 
(\ref{matrix}). 
This program reflects in its main features
the TRANSPORT beam optics program \cite{transport}
and is similar to the simulation program used to design
the Leading Proton Spectrometer at ZEUS \cite{LPS,transplps}.
The beta value at the intersection point considered in the calculation 
is 1000 m.

A short description of the main characteristics of the simulation
program is given in the following.
Given the beam momentum, a track is followed from the
starting point until it is lost at an obstruction.
The program returns the hit pattern at selected positions where detectors
may be placed.
The elements along the beam line which have been simulated are: 
\begin{itemize}
\item 
Drifts - in which no particle is lost and which have the same aperture 
as that of the previous element.
\item 
Quadrupoles - the focussing effect is calculated as a function of $x_L$.
Negative focal length means horizontal focussing.
Negative $x_{L}$ means particle with negative charge.

\item
Dipoles - the bending power is calculated with respect to that for a track 
at $x_{L} =1$, so that no bending occurs for particles 
at $x_{L} =1$.
During tracking inside a magnetic element the  
distance of the particle from the nominal beam may reach
a local maximum and a test is done to see if 
this maximum is outside the beam pipe.
The same test is performed at the entrance and at the exit of each quadrupole.

\item
Detectors - at present these are considered to be virtual planes
covering the whole beam pipe aperture except for a rectangular hole
with sides of length equal to $10\sigma_x $ and $10\sigma_y $,
where $\sigma_x $ and $\sigma_y $ are the horizontal and vertical 
standard deviation of the beam distribution 
(taken to be a Gaussian), respectively. 

\end{itemize}

For the following study particles were generated in the interaction 
vertex
diamond of dimensions $(\Delta x,\Delta y, \Delta z)\;=$(0.6mm, 0.6mm, 0.0mm),
according to a flat distribution in $log(1-x_L)$
within the limits $10^{-4} < (1-x_L) < 10^{-1}$ and a flat distribution 
in $log(p_T)$ within the limits $10^{-4} < p_T < 1$ MeV/c. The $z$ axis 
points in the proton beam direction.
The $x_L$ distribution has been chosen because it reflects
the observed single diffractive cross section, the $p_T$ distribution 
in order to 
enhance the generation of particles at very low $p_T$ 
(and thus at very low $|t|$)
and therefore approach the optical point
$(t=0)$ in the study of the elastic cross section at $(1-x_L) = 0$.

\vspace{1.cm}

\indent
\section{\bf The spectrometers }

\vspace{0.5cm}
\indent

In order to maximize the geometric acceptance
all the detectors are placed 
in the region beyond 200 m from the IP where the beam
profile 
has the smallest transverse size (see Fig.\ref{Fig:fig3}).

In this way we select
three regions where the beam elements 
can be used to momentum analyse very forward outgoing protons.
We call them spectrometers $A$, $B$, $C$. 
Each of them consists of two detector planes 
positioned upstream and downstream of a dipole element.
The detector planes are located only where the distance between
adjacent magnetic elements is $\gsim$ 50 cm. 

\begin{figure}[p]
\begin{picture}(450,550)(0,100)
\includegraphics{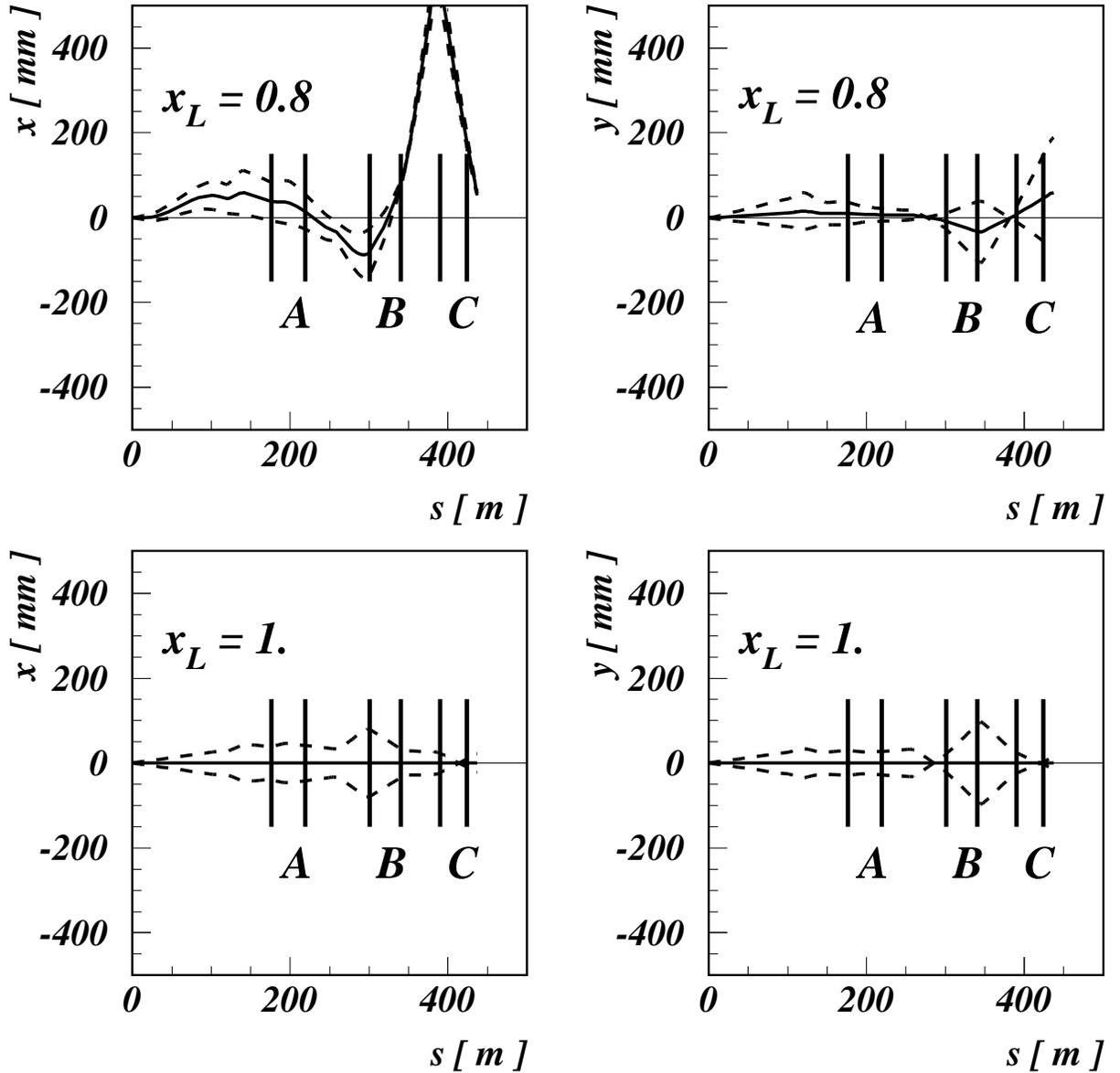}
\end{picture}
\vspace{-1.cm}
\caption{Typical horizontal and vertical trajectories of particles
through the magnetic elements of the beam line 
at different $x_L$ and $p_T$ values. The continuous lines 
correspond to particles with $p_T = 0$, the two dashed lines
correspond to particles with $p_T = \pm 2$ GeV/c. }
\label{Fig:fig4}
\end{figure}

Fig.\ref{Fig:fig4} shows particle trajectories through one of the 
outgoing arms from IP4.
They are calculated considering two values of $x_{L}$ and three values
of $p_T$. We choose two extreme $p_T$ values of $\pm 2\ \mbox{GeV/c}$  
and the central value $p_T=0$.
The beam pipe diameter has been taken to be infinite when tracking 
particles at this stage.
Vertical bands are superimposed to indicate the positions
of the three spectrometers.

Table \ref{dete} summarizes the 
positions and characteristics of the planes of each of the spectrometers.

In addition to spectrometers $A$, $B$, $C$ which cover the 
$very\ forward$
region, there is a 45 m long drift space downstream of the
central experimental region between dipoles $D1$ and $D2$ which
allows the detection
of the forward going leading particles with low $x_L$ down to $x_L \sim 0$.
The geometrical acceptance is calculated also for this region
in order to see which range of $x_L$ and $p_T$ can be detected outside the 
vacuum beam pipe.
A preliminary study of this region is given in \cite{Zichi}.

\vspace{1.cm}

\indent
\section{\bf Geometrical acceptance and momentum resolution}

\vspace{0.5cm}
\indent

At least three points are needed to measure the momentum.
We use the interaction vertex and demand the coincidence of any two 
detector planes (double coincidence events).
Using the transport matrix (\ref{matrix}) we relate the impact
point of the track on the detector
to the position and angle of the track at 
the interaction point.
At plane $i$ we can write the equation
       \begin{equation}
         x_i\;=\;H_{i,11}\,x(0)\,+\,H_{i,12}\,\theta_{x}(0)
       \label{rectra}.
       \end{equation}
\noindent
Assuming the nominal vertex ($x(0)$=0) we can also 
relate the position of the track at two different planes $i,j$
with the equation

       \begin{equation}
         x_i\;=\;M_{ij}(x_L)\,x_j\;+\;C_{ij}(x_L).
       \label{corre}
       \end{equation}

For a given pair of detectors $i,j$ 
and for a given value of $x_{L}$,
the coordinates $x_i$ and $x_j$ are thus linearly correlated;
the parameters of the straight line are functions of $x_L$ only,
thereby allowing the determination of $x_L$ from double coincidence
events. 

This method is used for momentum reconstruction in the ZEUS LPS 
\cite{LPS}
and a longitudinal momentum resolution 
of $0.4 \%$ has been achieved for particles
at $x_L=1$; the resolution on the transverse plane for the ZEUS LPS is 
less than the beam transverse momentum spread of $\sim 40 - 100$ MeV/c.

\begin{figure}[p]
\begin{picture}(450,550)(0,100)
\includegraphics{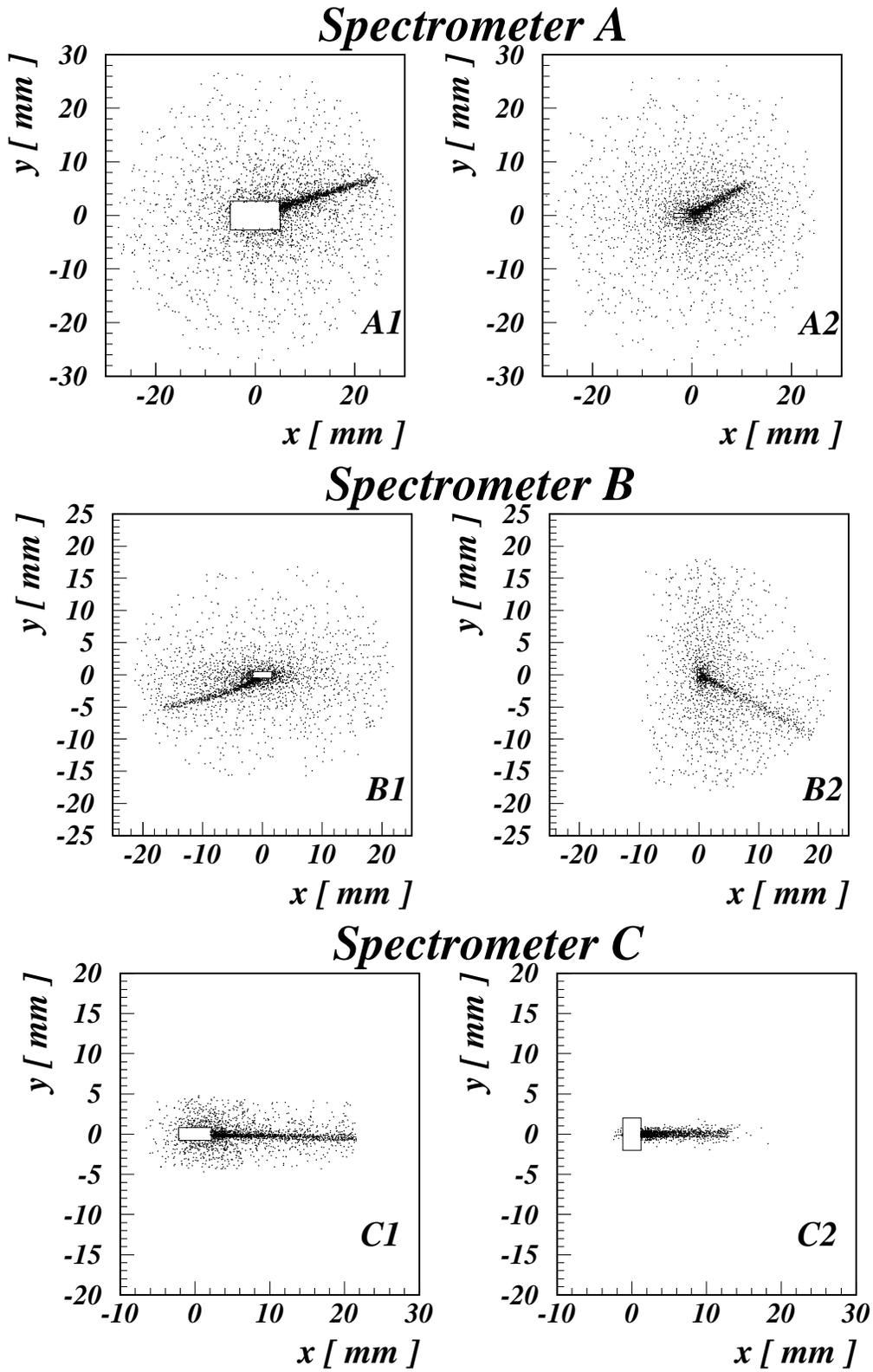}
\end{picture}
\vspace{1.cm}
\caption{Distribution of the hits at the detector planes of 
spectrometers $\bf A$, $\bf B$ and $\bf C$. A box is drawn at each plane
representing the dimensions of the $10 \sigma$ profile
of the beam.}
\label{Fig:fig5}
\end{figure}

Fig.\ref{Fig:fig5}
shows the pattern of hits at the detector planes
of the three spectrometers. The  
$10\sigma $ profile of the beam is often clearly visible.
The area covered by the hits on each plane is always asymmetric.
This asymmetry originates from the bending and focalizing
power of all the magnetic elements of the proton beam line
upstream of the detector planes; 
this must be taken into account in order to define the best geometry
of the detector planes.

\begin{figure}[p]
\begin{picture}(450,550)(0,100)
\includegraphics{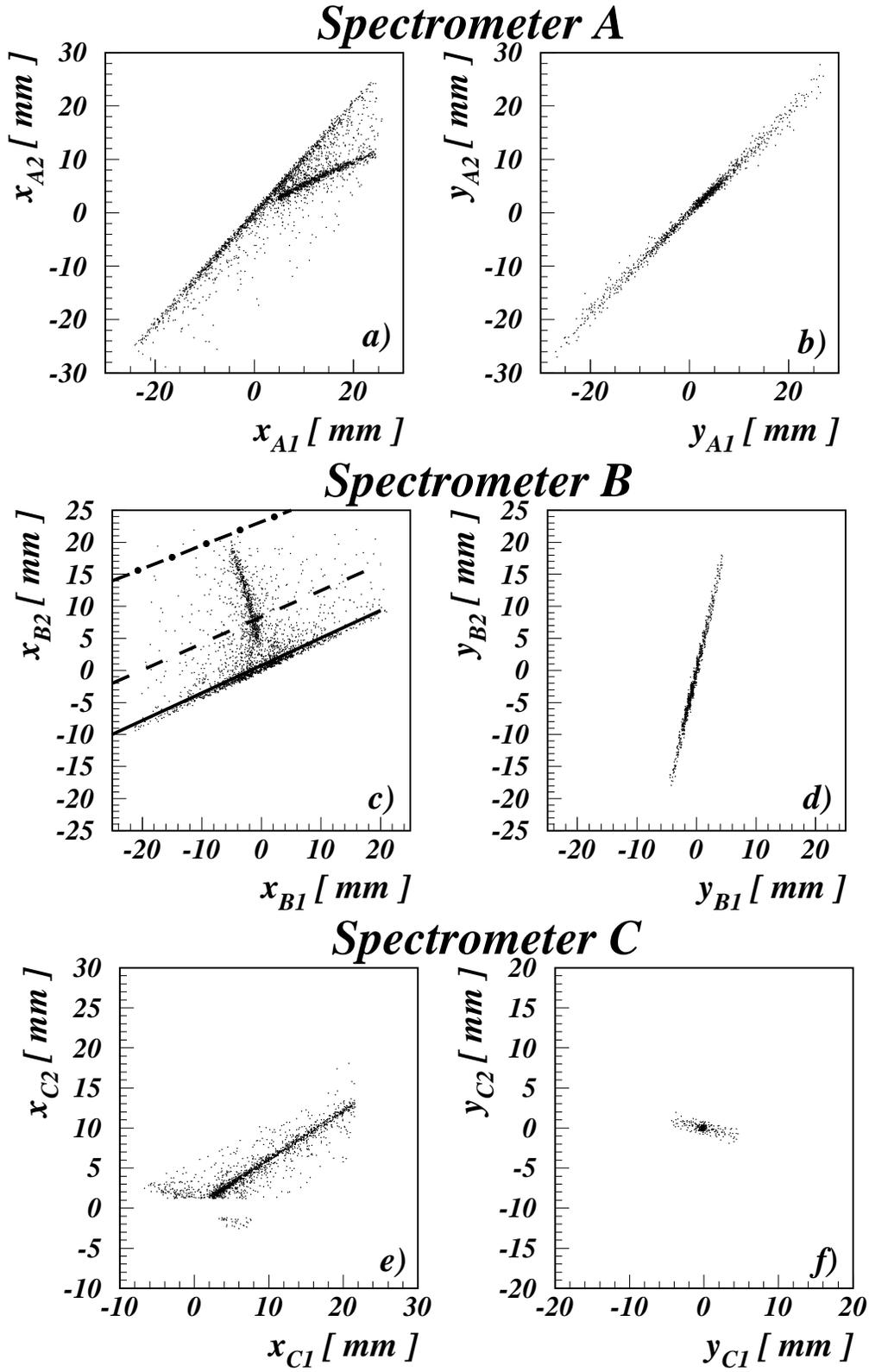}
\end{picture}
\vspace{1.cm}
\caption{Correlations between track positions in pairs of planes of the
spectrometers. Continuous, dashed, dot-dashed lines in (c)
correspond to tracks with $x_L=0.999$,$x_L=0.980$,$x_L=0.945$, respectively.}
\label{Fig:fig6}
\end{figure}

Fig.\ref{Fig:fig6} shows the correlation between horizontal 
and vertical coordinates
for the different pairs of detector planes which
are used to calculate the momentum.
The regions of the planes which detect very high $x_L$ and very small $p_T$
tracks correspond, in the figure, to the areas of higher event density. 
This is because of the particular 
generation we have used which is mentioned in the previous section.  
The straight lines, superimposed to the scatter plot 
shown in Fig.\ref{Fig:fig6}c, 
correspond to the different values of $x_{L}$ at which particles were
generated.
Correlations on the horizontal plane provide
a good $x_{L}$ resolution because the lines which 
represent two adjacent values of $x_{L}$ are well separated.
On the other hand on the vertical plane
the lines representing the same values of $x_L$
would be densely packed and are not used for resolution calculation. 

\begin{figure}[p]
\begin{picture}(450,550)(0,100)
\includegraphics{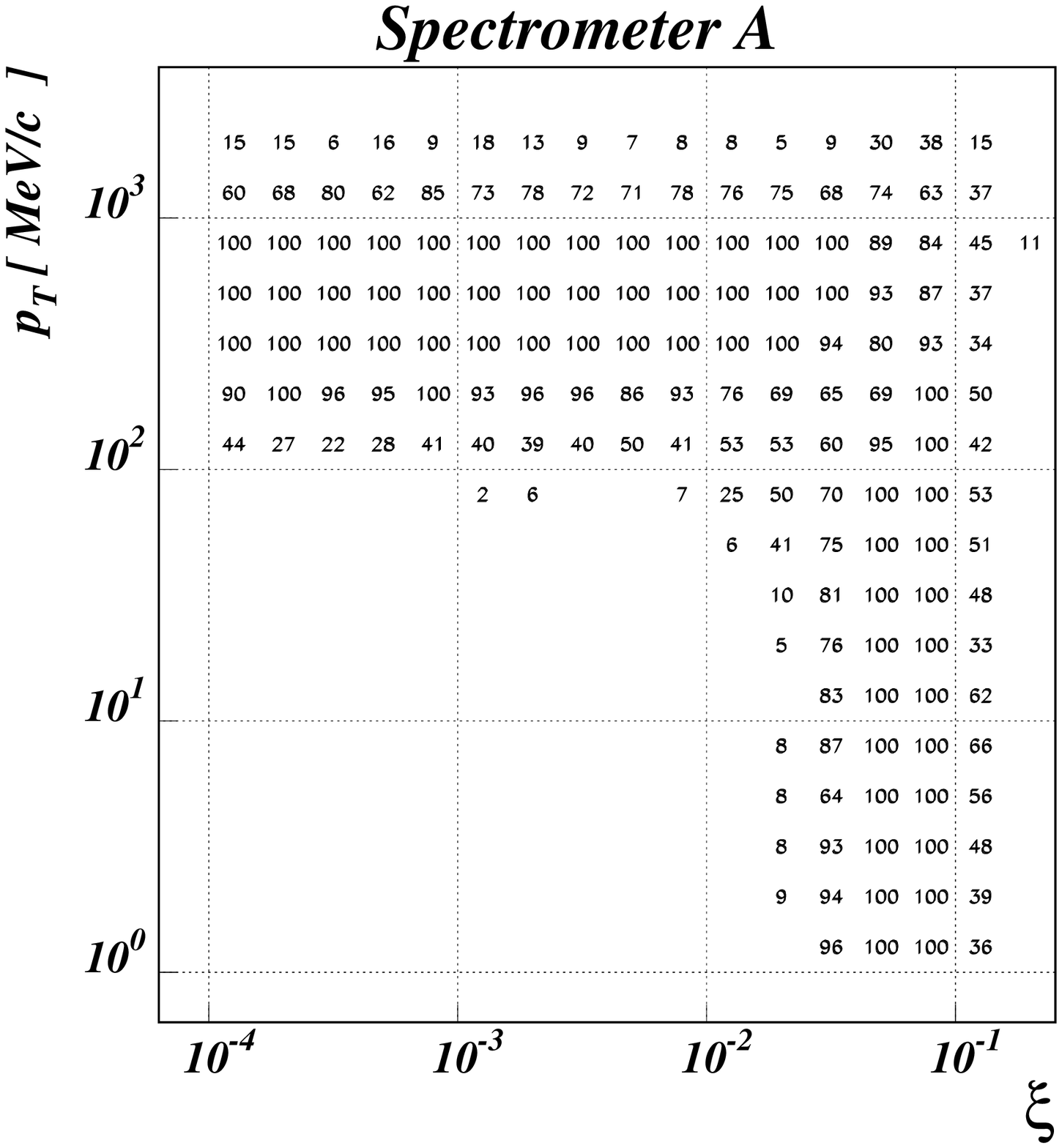}
\end{picture}
\caption{Geometrical acceptance as a function of $\xi$ and $p_T$
for double coincidence events 
in spectrometer $A$.}
\label{Fig:fig7}
\end{figure}

\begin{figure}[p]
\begin{picture}(450,550)(0,100)
\includegraphics{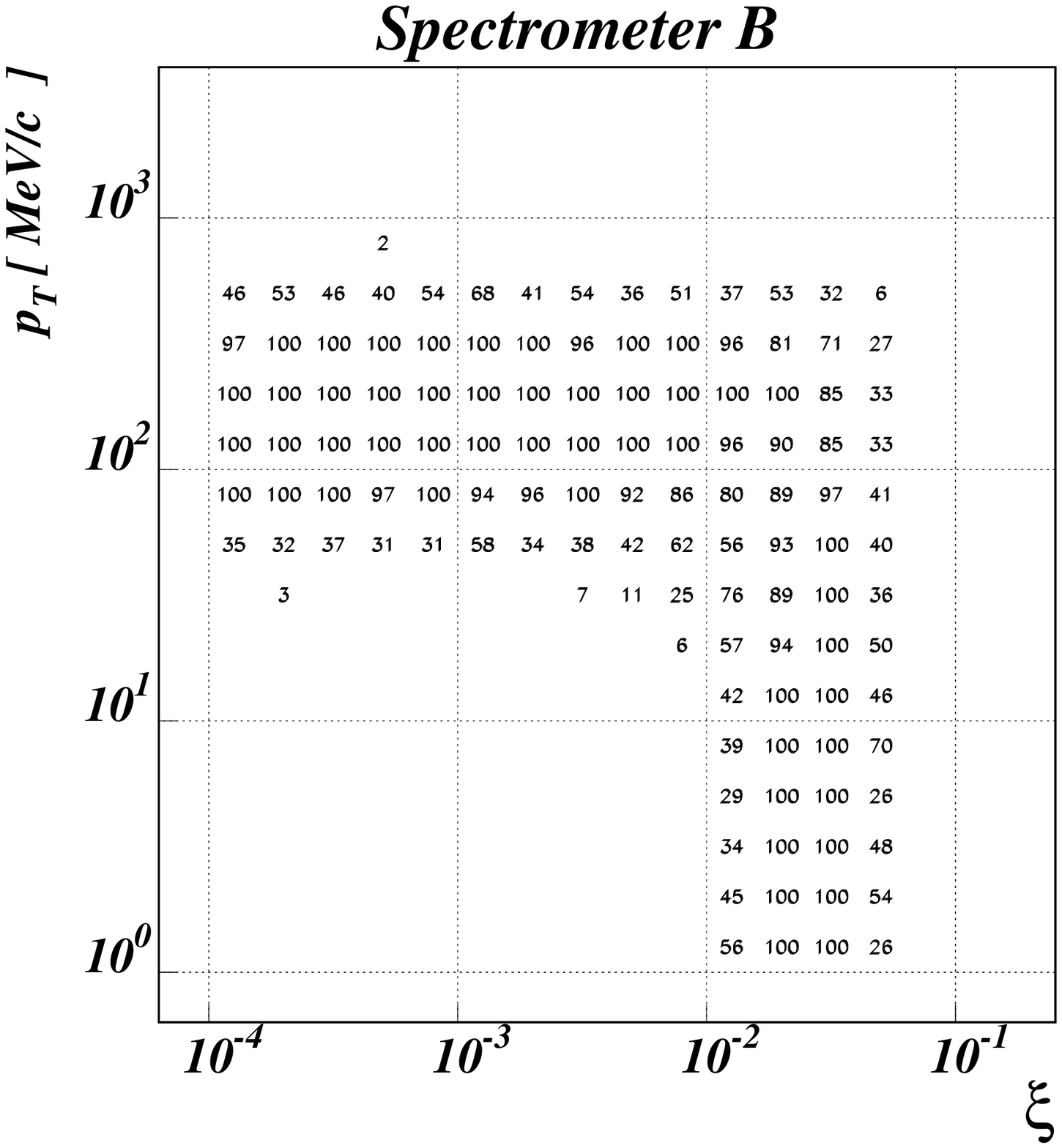}
\end{picture}
\caption{Geometrical acceptance as a function of $\xi$ and $p_T$
for double coincidence events 
in spectrometer $B$.}
\label{Fig:fig8}
\end{figure}

\begin{figure}[p]
\begin{picture}(450,550)(0,100)
\includegraphics{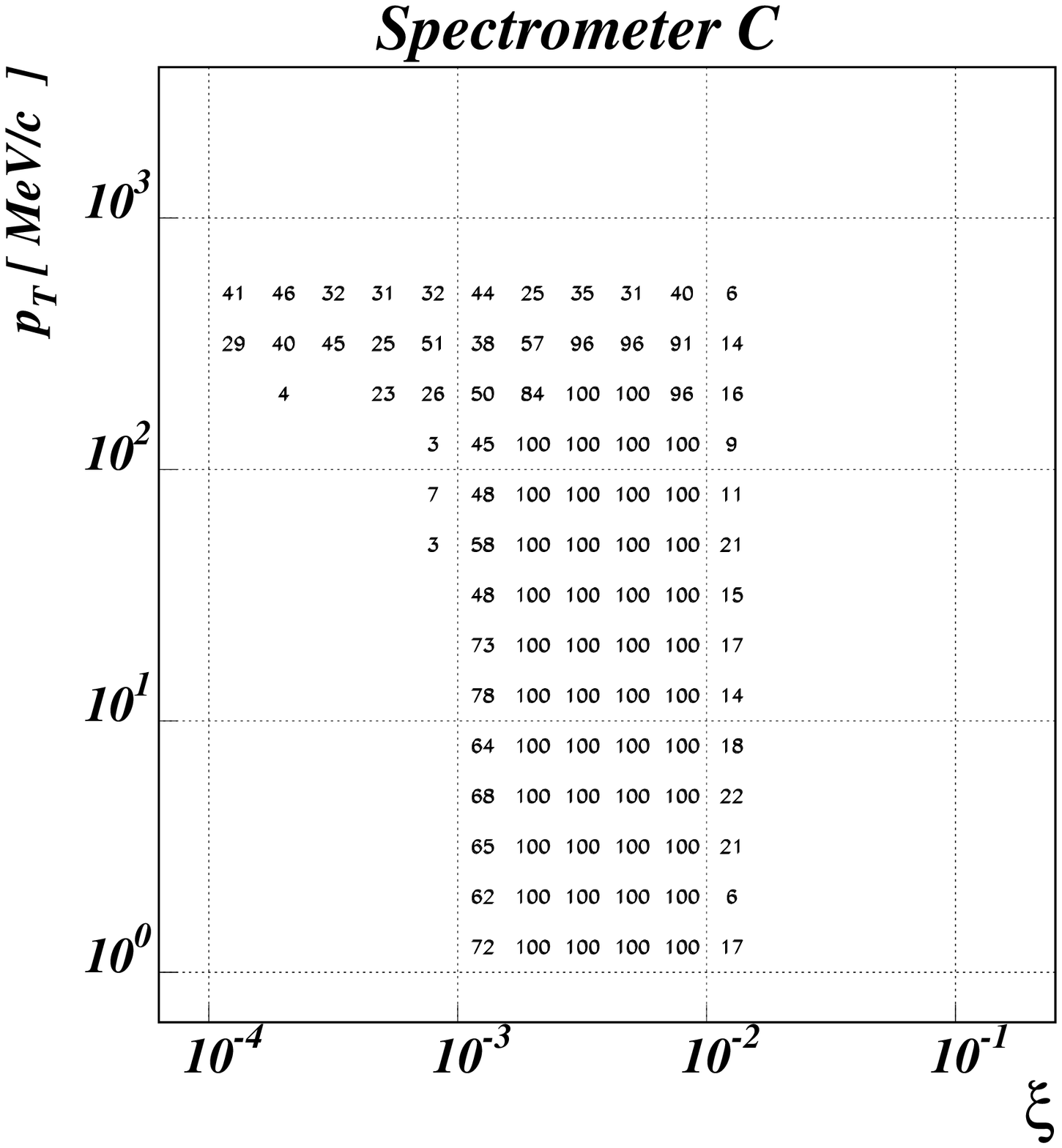}
\end{picture}
\caption{Geometrical acceptance as a function of $\xi$ and $p_T$
for double coincidence events 
in spectrometer $C$.}
\label{Fig:fig9}
\end{figure}

The geometric acceptance for the double coincidence events
is shown in Fig.\ref{Fig:fig7}, Fig.\ref{Fig:fig8} and
Fig.\ref{Fig:fig9} as a function of $\xi$ and $p_T$
for the spectrometers $A$, $B$ and $C$, respectively.
The acceptance is limited by the vacuum beam pipe apertures at the 
highest values of $\xi$ and the highest values
of $p_T$. On the other hand 
particles with very small $p_T$ and with momentum
close to the beam momentum 
escape within the $10 \sigma $ detector apertures.

From Fig.\ref{Fig:fig7}, Fig.\ref{Fig:fig8} and
Fig.\ref{Fig:fig9} we can conclude that spectrometer $A$
has $100 \%$ acceptance in the region $0.03 \lsim \xi \lsim 0.1$
and $1 \lsim p_T \lsim 10^3$ MeV/c and in the region
$10^{-4} \lsim \xi \lsim 0.1$ and $10^2 \lsim p_T \lsim 10^3$ MeV/c.
Spectrometer $B$ has $100 \%$ acceptance in the region
$10^{-2} \lsim \xi \lsim 0.03$ and $1 \lsim p_T \lsim 10^3$ MeV/c
and in the region $10^{-4} \lsim \xi \lsim 0.03 $ 
and $50 \lsim p_T \lsim 500$ MeV/c.
Finally spectrometer $C$ has $100 \%$ acceptance in the region
$10^{-3} \lsim \xi \lsim 10^{-2}$ and $1 \lsim p_T \lsim 10^3$ MeV/c.
In general very low $\xi$ tracks are accepted for
$10^2 \lsim p_T \lsim 10^3$ MeV/c, and very low $p_T$ tracks are accepted in 
the region $10^{-3} \lsim \xi \lsim 0.1$.

\begin{figure}[p]
\begin{picture}(450,550)(0,100)
\includegraphics{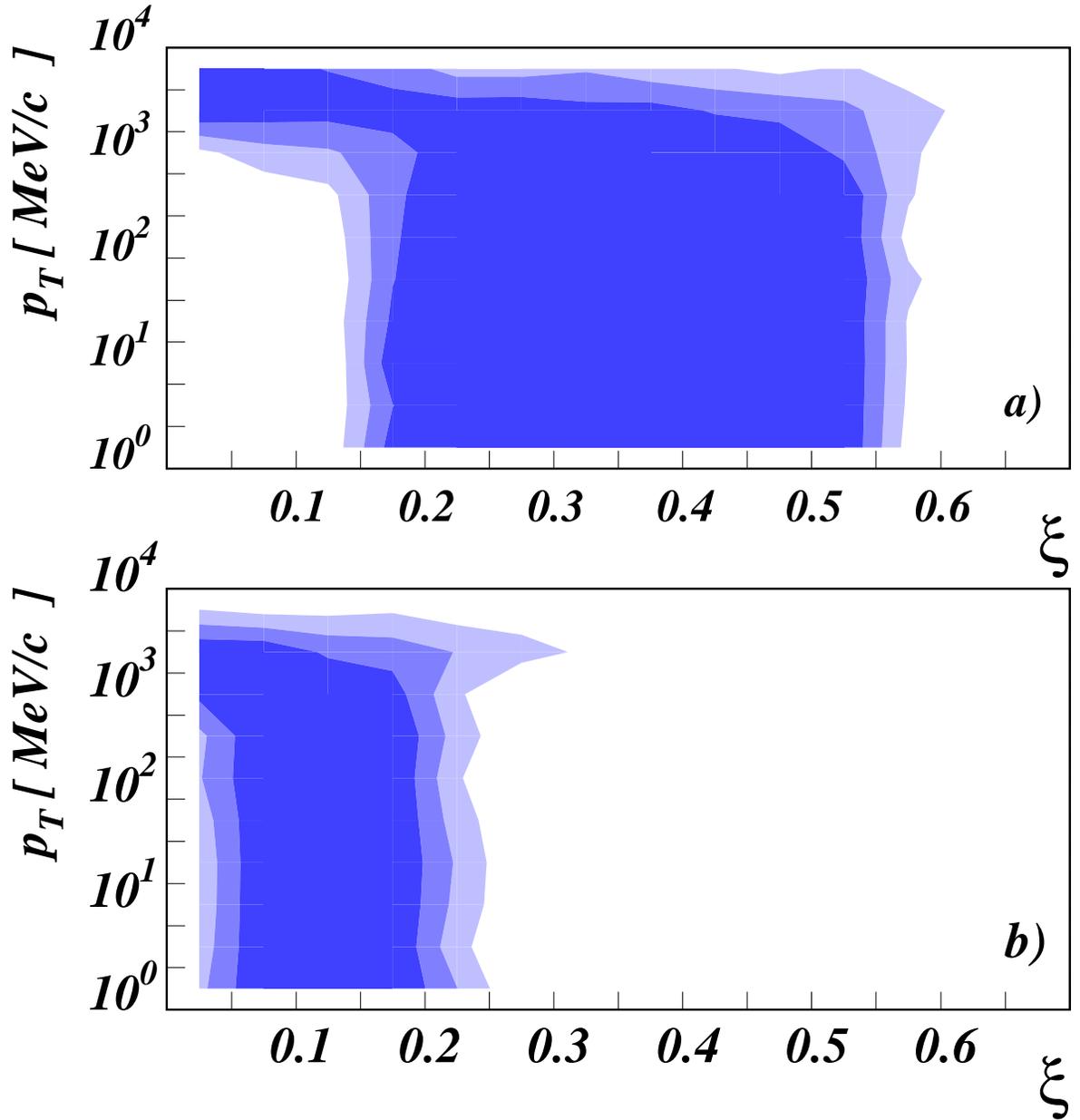}
\end{picture}
\caption{Geometric acceptance as a function of $\xi$ and $p_T$
for two planes placed at the beginning (a) and at the end (b)
of the 45 m drift space between dipoles D1 and D2.
Three acceptance ranges are shown in the pictures:
$\sim 100 \%$ (dark grey), $100-50 \%$, and $50-0 \%$ (light grey).}
\label{Fig:fig10}
\end{figure}

A different region of $\xi$ can be studied by 
a spectrometer placed in the 45 m long drift region between dipoles D1 and D2.
Two detector planes have been assumed at the beginning 
and at the end of this drift space
and their geometrical acceptance is shown in 
Fig.\ref{Fig:fig10}
as a function of $\xi$ and $p_T$.
In this case $\xi$ has been generated uniformly between 0 and 1.
The figure shows a wide interval of $\xi$ between 
0.15 and 0.6 in which the acceptance at the beginning of the drift space
is $100 \%$ for particles inside the pipe. The acceptance does
not depend on $p_T$.
At the end of the drift space only particles
with $\xi \lsim 0.2$ stay inside the beam pipe
and, if we require the coincidence of the two planes, 
particles are accepted only in a very limited region of $\xi$.
This means that particles with $0.2 \lsim \xi \lsim 0.6$
can be analysed with detectors (and magnets)
placed outside the vacuum pipe.

\begin{figure}[p]
\begin{picture}(450,550)(0,100)
\includegraphics{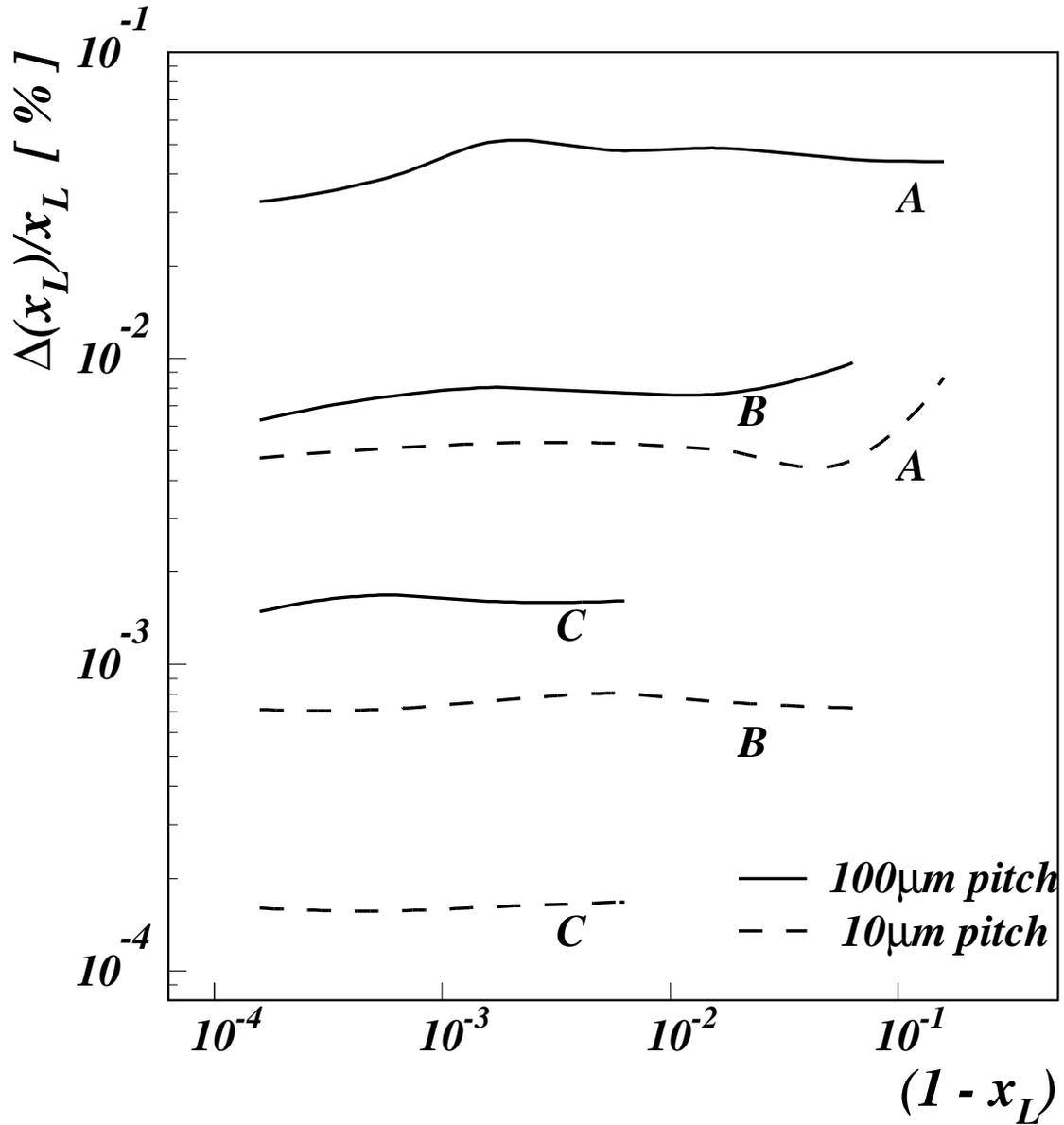}
\end{picture}
\caption{Momentum  resolution ${\Delta x_L} \over x_L$
for the spectrometers $A$, $B$ and $C$. 
Results for $100 \mu$m (continuous lines) 
and $10 \mu$m (dashed lines) pitch detectors are shown. }
\label{Fig:fig11}
\end{figure}

The momentum resolution ${\Delta x_L} \over x_L$ has been 
calculated for the three spectrometers 
with either $100 \mu$m or $10 \mu$m pitch detectors
and is shown in Fig.\ref{Fig:fig11}.
Two different sets of curves are shown in the figure. The continuous
lines show the momentum resolution for the three spectrometers
$A, B, C$ equipped with $100 \mu$m pitch detectors, while the dashed lines
show the resolutions for $10 \mu$m pitch detectors.
All curves show an approximately constant behaviour
in the considered $(1-x_L)$ interval, and in general the spectrometers
with $10 \mu$m pitch detectors show a resolution which is one order of
magnitude better than the stations with $100 \mu$m pitch detectors.
The resolution improves as the distance from the interaction point
increases. This effect originates 
from the increased integrated magnetic field
acting on the particle.
The best resolution is obtained with spectrometer C and has the values 
${{\Delta x_L} \over x_L} \sim 10^{-3} \%$ and
${{\Delta x_L} \over x_L} \sim 10^{-4} \%$ for the 
$100 \mu$m and $10 \mu$m pitch detectors, respectively.
Spectrometer C accepts however only particles with $(1-x_L) \lsim 10^{-2}$
due to the large distance of the detectors from the interaction point.
We can conclude that the order of magnitude of the momentum resolution
is always better of 1 GeV/c for a typical momentum of $7 \cdot 10^3$ GeV/c.

The momentum resolution calculated using vertical coordinate 
correlations is very poor compared to the previous one.
This can be seen also from Fig.\ref{Fig:fig6} 
where the vertical correlation lines are very densely packed
and generate many ambiguities in the momentum calculation. 

The transverse momentum resolution 
 ${{\Delta p_T}\over p_T}={{({p_T}(rec)-{p_T}(gen))}\over {p_T}(gen)}$ 
as a function of ${p_T}(gen)$ has been studied for the three spectrometers
separately, with either $100 \mu$m or $10 \mu$m pitch detectors.

The transverse momentum $p_T$ has been calculated using
the two components $p_x$ and $p_y$ 
computed using the relation:
\begin{equation}
         p_x(0)\;=\;\theta_{x}(0)\,x_L(hor)\,p_{beam},
       \label{px}
\end{equation}
where $\theta_{x}(0)$ is obtained from the transport matrix
eq.(\ref{matrix}), and $x_L(hor)$ is the momentum estimate
in the horizontal plane.
The $p_y$ component is calculated using a
similar equation in the $y$ plane, but still
considering the momentum estimate  $x_L(hor)$.

\begin{figure}[p]
\begin{picture}(450,550)(0,100)
\includegraphics{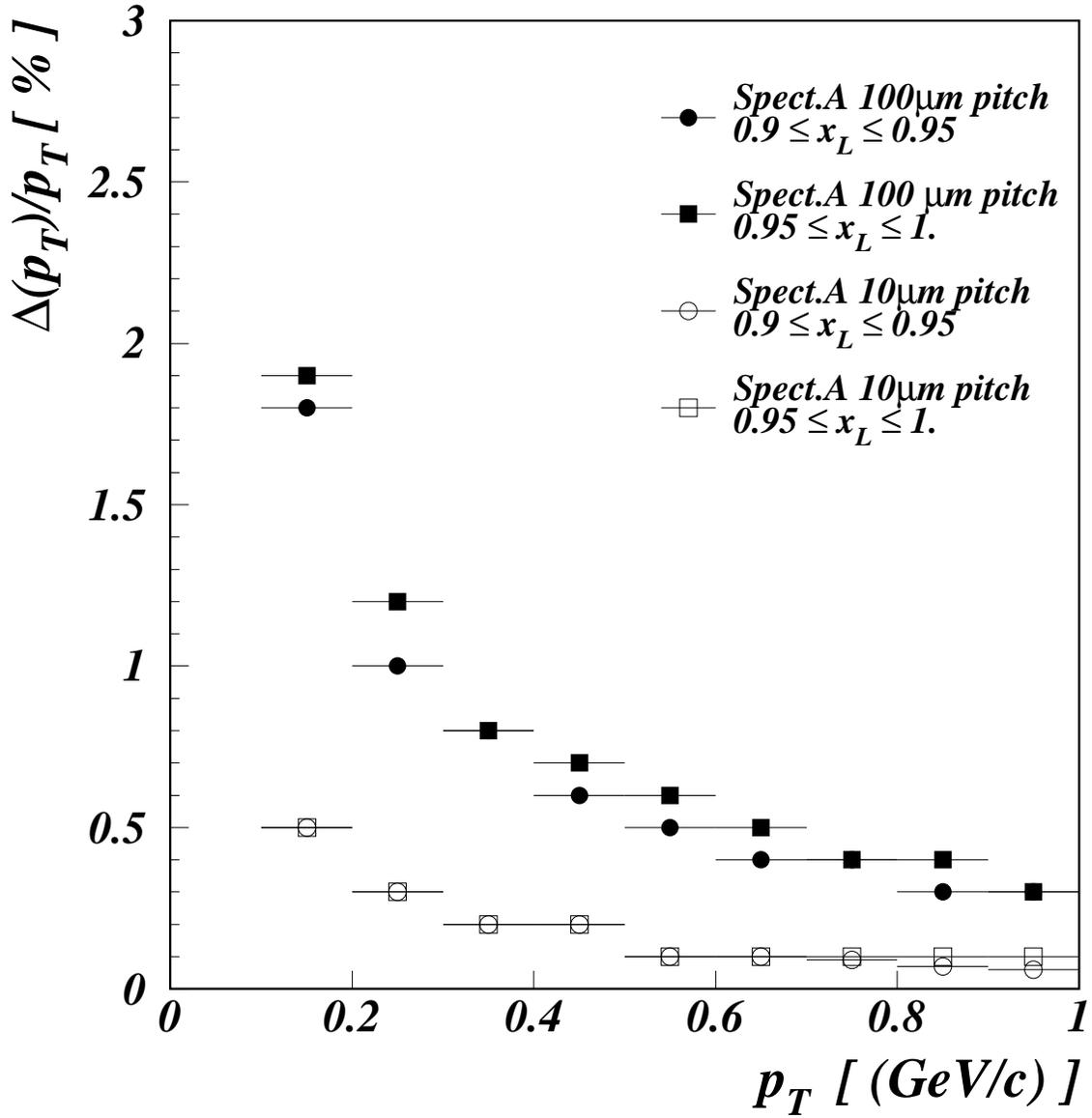}
\end{picture}
\caption{Transverse momentum  resolution
 ${{\Delta p_T}\over p_T}$ 
for spectrometer $A$ for particles generated
in two different $x_L$ bins and for $100 \mu$m (full symbols) or 
$10 \mu$m (open symbols) pitch detectors.}
\label{Fig:fig12}
\end{figure}

\begin{figure}[p]
\begin{picture}(450,550)(0,100)
\includegraphics{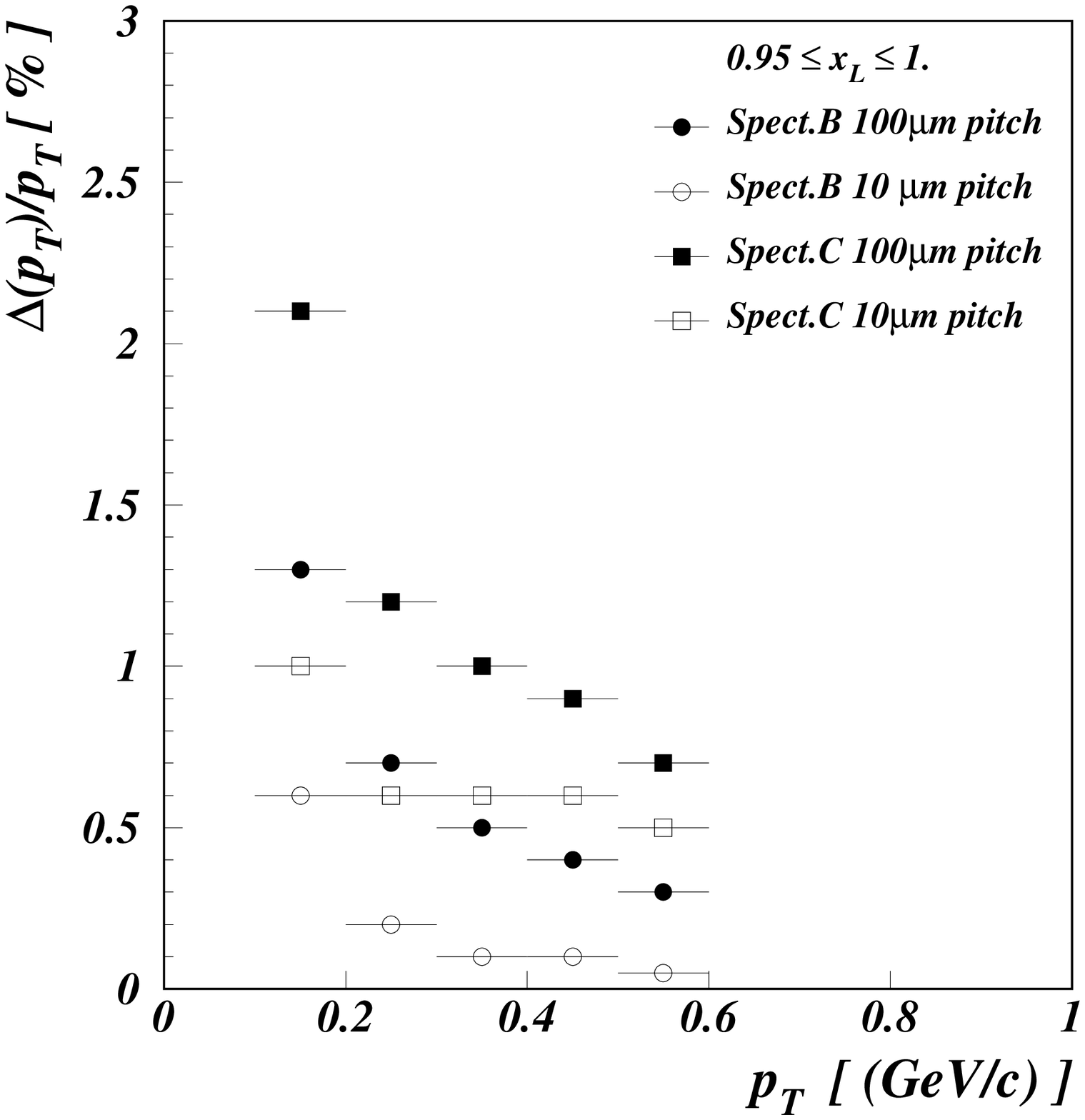}
\end{picture}
\caption{Transverse momentum  resolution
 ${{\Delta p_T}\over p_T}$ 
for spectrometers $B$  and $C$ equipped with
$100 \mu$m (full symbols) or 
$10 \mu$m (open symbols) pitch detectors.}
\label{Fig:fig13}
\end{figure}

Fig.\ref{Fig:fig12} shows the $p_T$ resolution for spectrometer A 
for two different detector pitches.
We can observe that the resolution is always better that $3 \%$ for 
$p_T \gsim 200$ MeV/c and slowly improves as a function of $p_T$.
No $x_L$ dependence is observed both in the case of
$100 \mu$m and  $10 \mu$m pitch detectors.
The resolution is a factor of two better for the
$10 \mu$m pitch detectors planes in all $p_T$ bins considered.
The best resolution in spectrometer A is of the order of $0.1 \%$.
Fig.\ref{Fig:fig13} shows the $p_T$ resolution for the spectromenter 
B and C.
Full and open symbols represent the resolution for the $100 \mu$m
and for the $10 \mu$m pitch detectors, respectively.
In all cases the resolution slowly improves 
as a function of $p_T$ and takes values which are approximately
of the same order of magnitude as those observed in spectrometer A.
Finally it is interesting to notice that the calculated resolution
in the whole $p_T$ range considered is of the
order of the nominal beam transverse momentum spread ($\lsim 5$ MeV/c).

\vspace{1.cm}

\indent

\section{\bf Conclusions}
 
\vspace{0.5cm}

\indent

Geometrical acceptance and momentum resolution
have been calculated for three spectrometers along one of the
outgoing proton beam arms of IP4 at LHC
in the interval $10^{-4} \lsim \xi \lsim 0.1$
and $0 \lsim p_T \lsim 10^3$ MeV/c.
The apparatus has been simulated with 
$100 \mu$m or with $10 \mu$m pitch detector planes.
We have shown that the three spectrometers cover regions of $x_L$
which partially overlap. They have very different momentum
resolutions for equal values of $x_L$ but their $p_T$ resolution
do not differ dramatically.
In general going from $100 \mu$m to $10 \mu$m pitch 
an order of magnitude in the momentum resolution is gained,
while only a factor two is gained for the $p_T$ resolution.
The momentum resolution is in the range
$10^{-4} \% \lsim {{\Delta x_L} \over x_L} \lsim 5 \cdot 10^{-2} \%$.
The $p_T$ resolution is in the range
$0.1 \% \lsim {{\Delta p_T} \over p_T} \lsim 3 \%$ for particles with
$p_T \gsim 200$ MeV/c.

\vspace{1.cm}

\indent

{\bf Aknowledgments}
 
\vspace{0.5cm}

\indent

The results presented in this paper were obtained in the early stages
of the effort that recently led to the completion of the FELIX Letter
of Intent \cite{felix2}. We are grateful to the authors of \cite{felix1}
for many discussions. In particular we are indebted to
K. Eggert and A. Morsch for their interest in our work, for many
useful suggestions and for carefully reading the manuscript.

\vspace{0.5cm}
\indent


\clearpage

\begin{table}[p]
\begin{center}

\begin{tabular}{|c|c|c|c|c|c|} \hline
     &      &       &        &          &           \\
Name & Type & start & length & spot considered& parameter \\
     &      &       &        &         &    \\
     &      &[m]    &[m]     &[mm] x [mm] &[mrad] if bend     \\
     &      &       &        &          &  [m] if quad         \\
     &      &       &        &          &           \\ \hline
     &      &       &        &          &           \\
LALEPH   & drift& 0.& 3.83 & - &  \\
DUA1& bend & 3.83& 4.95 & - & 0.170 \\
L2  & drift & 8.78 & 2.72 & - & \\
D0  & bend & 11.5 & 2.0 & - & 0.342 \\
L3  & drift & 13.5 & 4.5 & - & \\
D1  & bend & 18.0 & 9.45 &44 x 44 & 1.62 \\
L4  & drift& 27.45 & 0.6 & &  \\
D1  & bend & 28.05 & 9.45 & 44 x 44& 1.62 \\
L5  & drift & 37.5 & 45.265 & & \\
D2  & bend & 82.765 & 9.45 &44 x 44 & -1.62 \\
L6  & drift & 92.215 & 0.6 & & \\
D2  & bend & 92.815 & 9.45 &44 x 44 & -1.62 \\
L7  & drift & 102.265 & 2.2 & & \\
Q1  & quad & 104.465 & 5.5 &35 x 35 & 74.822 \\
L8  & drift & 109.965 & 4.5 & & \\
Q2  & quad & 114.465 & 5.5 &35 x 35 & -52.397 \\
L9  & drift & 119.965 & 1.0 & &  \\
Q2  & quad & 120.965 & 5.5 &35 x 35 & -52.397 \\
L10 & drift & 126.465 & 5.5 & &  \\
Q3  & quad &  131.965 & 5.5 &35 x 35 & 39.440 \\
L11 & drift & 137.465 & 1.8 & & \\
    &      &         &     & &           \\
    &      &         &     & &           \\ \hline
\end{tabular}
\end{center}
\caption{Positions of the beam elements 
of one of the two outgoing proton beams
with respect to the IP4
and the corresponding field values
considered in the simulation program. 
The spot considered 
is the area transverse to the nominal beam
available for the passage of particles.}
\label{geo1}
\end{table}

\clearpage

\begin{table}[p]
\begin{center}

\begin{tabular}{|c|c|c|c|c|c|} \hline
     &      &       &        &          &           \\
Name & Type & start & length & spot considered& parameter \\
     &      &       &        &       &    \\
     &      &[m]    &[m]     &[mm] x [mm]     &[mrad] if bend  \\
     &      &       &        &          &    [m] if quad         \\
     &      &       &        &          &           \\ \hline
     &      &       &        &          &           \\
L12 & RF cav& 139.265 & 33.528 & & \\
Q4  & quad & 172.793 & 3.0 &28 x 28 & -123.001\\
L13 & drift & 175.793 & 20.0 & & \\
Q5  & quad & 195.793 & 3.0 &28 x 28 & 85.251\\
L14 & drift & 198.793 & 1.2 & &  \\
D3  & bend  & 199.993 & 9.45 &38 x.38 & -1.62 \\
L15 & drift & 209.443 & 0.6 & & \\
D3  & bend  & 210.043 & 9.45 &38 x.38 & -1.62 \\
L16 & drift & 219.493 & 16.0 & & \\
D4  & bend  & 235.493 & 9.45 &38 x.38 & 1.62 \\
L17 & drift & 244.943 & 0.6 & & \\
D4  & bend  & 245.543 & 9.45 &38 x.38 & 1.62 \\
L18 & drift & 254.993 & 1.0 & & \\
L19 & drift & 255.993 & 0.2 & & \\
Q6  & quad & 256.193 & 3.0 &28 x 28 & -45.045 \\
L20 & drift & 259.193 & 0.2 & & \\
Q6  & quad & 259.393 & 3.0 &28 x 28 & -45.045 \\
L21 & drift & 262.393 & 2.5 & & \\
BDS0& bend & 264.893 & 14.2 &  & 5.1 \\
L22 & drift & 279.093 & 1.46 & & \\
BDS0& bend & 280.553 & 14.2 &  & 5.1 \\
L23 & drift & 294.753 & 2.14 & & \\
Q7  & quad & 296.893 & 3.25 &22 x 18 & 34.150 \\
L24 & drift & 300.143& 0.2 & & \\
Q7  & quad & 300.343 & 1.5 &22 x 18 & 1536.098 \\
    &      &         &     & &           \\
    &      &         &     & &           \\ \hline
\end{tabular}
\end{center}
\caption{Same as Table 1. }
\label{geo2}
\end{table}

\begin{table}[p]
\begin{center}

\begin{tabular}{|c|c|c|c|c|c|} \hline
     &      &       &        &          &           \\
Name & Type & start & length & spot considered& parameter \\
     &      &       &        &          &    \\
     &      &[m]    &[m]     &[mm]      & [mrad] if bend  \\
     &      &       &        &          & [m] if quad  \\
     &      &       &        &          &           \\ \hline
     &      &       &        &          &           \\
L25 & drift & 301.843 & 8.91 & & \\
BDS0& bend & 310.753 & 14.2 & & 5.1 \\
L26 & drift & 324.953 & 1.46 & & \\
BDS0& bend & 326.413 & 14.2 & & 5.1 \\
L27 & drift & 340.613 & 2.14 & & \\
Q8  & quad & 342.753 & 3.25 &22 x 18 & -34.150 \\
L28 & drift & 346.003& 0.2 & & \\
Q8  & quad & 346.203 & 1.5 & 22 x 18& -139.762 \\
L29 & drift & 347.703 & 4.86 & & \\
BDS0& bend & 352.563 & 14.2 & & 5.1 \\
L30 & drift & 366.763 & 1.46 & & \\
BDS0& bend & 368.233 & 14.2 & & 5.1 \\
L31 & drift & 382.423 & 2.14 & & \\
Q9  & quad & 384.563 & 3.25 &22 x 18 & 34.150\\
L32 & drift & 387.813& 0.2 & & \\
Q9  & quad & 388.013 & 1.5 &22 x 18 & 175.438 \\
L33 & drift & 389.513 & 4.99 & & \\
BDS0& bend & 394.503 & 14.2 & & 5.1 \\
L34 & drift & 408.703 & 1.46 & & \\
BDS0& bend & 410.200 & 14.2 & & 5.1 \\
L35 & drift & 424.363 & 9.321 & & \\
Q10 & quad & 433.684 & 1.5 &22 x 18 & -959.233 \\
L36 & drift & 435.184 & 0.2 & & \\
Q10 & quad & 435.384 & 1.5 &22 x 18 & -73.992 \\
    &      &         &     & &           \\
    &      &         &     & &           \\ \hline
\end{tabular}
\end{center}
\caption{Same as Table 1. }
\label{geo3}
\end{table}

\clearpage

\begin{table}[p]
\begin{center}

\begin{tabular}{|c|c|c|c|} \hline
\multicolumn{4}{|c|}{ \bf Spectrometer A} \\ \hline
Detectors & Position & $\sigma_x$of beam & $\sigma_y$ of beam \\
          & [m]     &[mm]           &[mm]       \\ \hline
     A1 & 175.793 & 0.50 & 0.27\\
     A2 & 219.493 & 0.37 & 0.01\\
         &        &          &           \\ \hline
\hline
\multicolumn{4}{|c|}{ \bf Spectrometer B} \\ \hline
Detectors & Position & $\sigma_x$of beam & $\sigma_y$ of beam \\
          & [m]     &[mm]           &[mm]       \\ \hline
     B1 &  301.843& 0.17 &0.06 \\
     B2 &  340.613& 0.04 &0.04 \\
         &        &          &           \\ \hline
\hline
\multicolumn{4}{|c|}{ \bf Spectrometer C} \\ \hline
Detectors & Position & $\sigma_x$of beam & $\sigma_y$ of beam \\
          & [m]     &[mm]           &[mm]       \\ \hline
     C1 &  389.513&0.23  &0.09 \\
     C2 &  424.363&0.13  &0.21 \\
    &      &         &             \\ \hline
\end{tabular}
\end{center}
\caption{List of detector planes considered in the three 
spectrometers and $1\sigma$ beam sizes at the planes. }
\label{dete}
\end{table}

\clearpage

\clearpage
\clearpage

\clearpage

\clearpage

\clearpage

\clearpage

\clearpage

\clearpage

\clearpage

\clearpage

\clearpage

\clearpage

\clearpage


\begin{thebibliography}{99}

\bibitem{Pom}
See e.g.:\\
T.Regge, Nuovo Cimento 14~(1959)~951;\\
G.F.Chew, S.C. Frautschi, Phys., Rev. Lett. 7~(1961)~394;\\
A.H.Mueller, Phys Rev. D2~(1970)~2963;\\
F.E.Low, Phys. Rev. D12~(1975)~163;\\
S. Nussinov, Phys. Rev. D14~(1976)~246;\\
U. Amaldi, M. Jacob and G. Matthiae, Ann. Rev. Nucl. Sci. 26~(1976)~385;\\
P.D.B. Collins, ``An Introduction to Regge Theory and High Energy Physics'',
Cambridge University Press, Cambridge, 1977;\\
A.B. Kaidalov, Phys. Rep. 50~(1979)~157;\\
G.Alberi and G. Goggi, Phys. Rep.74~(1981)~1;\\
K. Goulianos, Phys. Rep. 101~(1983)~169;\\
L.V. Gribov, E.M. Levin and M.G. Ryskin, Phys. Rep. 100~(1983)~1;\\
N.Arteaga-Romero et al., Mod. Phys. Lett. A1~(1986)~221;\\
G. Ingelman, P. Schlein, Phys. Lett. 152B~(1985)~256;\\
A. Donnachie, P.V. Landshoff, Nucl. Phys. B 303~(1988)~634;\\
N.N.Nikolaev, B.G.Zakharov, Z. Phys. C49~(1991)~607;\\
J.D. Bjorken, SSC EOI 19, Int. J. Mod.Phys. A7~(1992)~4189.
\bibitem{pots}
U.Amaldi et al., Phys. Lett. B 43~(1973)~231.
\bibitem{hadhad}
See e.g.:\\
Y.Akimov et al., Phys. Rev. D14~(1976)~3148;\\
M.G. Albrow et al., Nucl. Phys. B 108~(1976)~1;\\
R.L.Cool et al., Phys. Rev. Lett. 47~(1981)~701;\\
M.Bozzo et al., UA4 Collaboration, Phys. Lett. 136B~(1984)~217;\\
M. Bozzo et al., Phys. Lett. B 136~(1984)~217;\\
K.Eggert, UA1 Collaboration, Proceedings of the Int.
Conference on Elastic and Diffractive Scattering 15-18 Oct.1987,
Ed. K.Goulianos, The Rockefeller Univ., New York, p.1;\\
R. Bonino et al., UA8 Collaboration, Phys. Lett. B 211~(1988)~239;\\
A.Brandt et al., UA8 Collaboration, Phys. Lett. B 297~(1992)~417.
\bibitem{HERA}
See e.g.:  \\
ZEUS Collaboration, M. Derrick et al., Z. Phys. C68~(1995)~569;\\
H1 Collaboration, T. Ahmed et al., Phys. Lett. B348~(1995)~681;\\
H1 Collaboration, T. Ahmed et al., Nucl. Phys. B435~(1995)~3;\\
ZEUS Collaboration, M. Derrick et al., Phys. Lett. B346~(1995)~399;\\
ZEUS Collaboration, M. Derrick et al., Phys. Lett. B350~(1995)~120;\\
ZEUS Collaboration, M. Derrick et al., Phys. Lett. B356~(1995)~129;\\
ZEUS Collaboration, M. Derrick et al., Z. Phys. C69~(1995)~39;\\
ZEUS Collaboration, M. Derrick et al., Z. Phys. C70~(1996)~391;\\
H1 Collaboration, T. Ahmed et al., Z. Phys. C70~(1996)~609;\\
H1 Collaboration, T. Ahmed et al., Nucl. Phys. B472~(1996)~32;\\
ZEUS Collaboration, M. Derrick et al., Phys. Lett. B380~(1996)~220;\\
ZEUS Collaboration, M. Derrick et al., Z. Phys. C73~(1996)~73.
\bibitem{LPS}
ZEUS Collaboration, ``The ZEUS Detector, Status Report 1993", PRC/93;\\
ZEUS Collaboration, M. Derrick et al., accepted by Z. Phys. C73~(1997)~253.
\bibitem{diffLHC}
G. Matthiae, Proceedings Large Hadron Collider Workshop CERN 90-10, Vol 2;\\
TOTEM Collaboration, M.Bozzo et al., EOI CERN/LHCC 93-47;\\
K. Eggert and A. Morsch, CERN AT/94-09;\\
K. Eggert and A. Morsch, Nucl. Instr. and Meth. A351~(1994)~174.
\bibitem{Zichi}
T. Taylor, H.Wenninger and A. Zichichi, CERN-LAA/95-15.
\bibitem{felix1}
Y. Takahashi et al., CERN/LHCC 95-51 (LHC);\\
K. Eggert, A. Morsch and C. Taylor  ``A full acceptance
experiment at the CERN Large Hadron Collider (LHC)'' , VII th Blois workshop on
Elastic and Diffractive Scattering, June 1995 published in Frontiers in Strong
Interactions (1995), editors P. Chiapetta, M.Haguenauer and J. Tran Thanh Van;\\
K. Eggert and C. Taylor, "FELIX, A Full Acceptance Detector for the CERN LHC",
Nucl. Phys. B (Proc. Suppl.) 52B (1997) 279.
\bibitem{silvia}
S. Maselli, ``A small angle spectrometer for very forward physics at LHC''
Proceedings of the Sixth Topical Seminar on Experimental Apparatus
for Particle Physics and Astrophysics, S.Miniato, 20-24 May 1996,
Nucl. Phys. B54~(1997)~74.
\bibitem{MAD}
The LHC Study Group, CERN/AC/95-05 (LHC).\\
\bibitem{transport}
K.L.Brown et al., CERN Yellow Report 80-04.\\
\bibitem{transplps}
T. Massam, BEAM9, a program to simulate the Leading Proton Spectrometer 
at ZEUS, 1992 (unpublished).\\

At the time of printing the following references became avaliable:\\
\bibitem{felix2}
FELIX, ``A full acceptance detector at the LHC'', Letter-of-Intent LHCC/I10,
CERN/LHCC 97-45, 1st Aug. 1997.
\bibitem{beamline}
A. Verdier, ``A tunable insertion in point 4 for LHC'' LHC Project
Note 93, June 1997.

\end{thebibliography}
\end{document}